\documentclass[twocolumn,pra,showpacs,superscriptaddress,amssymb,amsmath,amsmath,floatfix]{revtex4-1}
\usepackage{graphicx}
\usepackage{epstopdf}
\usepackage{bm}
\usepackage[colorlinks=true,linkcolor=blue,urlcolor=blue,citecolor=blue]{hyperref}
\usepackage{comment}
\usepackage{float}
\usepackage{cancel}
\usepackage{times}
\usepackage{xcolor}

\newcommand{\be}{\begin{equation}}
\newcommand{\ee}{\end{equation}}

\begin{document}
\title{Interaction potentials, electric moments, polarizabilities, and chemical reactions \\ of YbCu, YbAg, and YbAu molecules}

\author{Micha\l~Tomza}
\email{michal.tomza@fuw.edu.pl}
\affiliation{Faculty of Physics, University of Warsaw, Pasteura 5, 02-093 Warsaw, Poland}

\date{\today}

\begin{abstract}

Ultracold YbAg molecules have been recently proposed as promising candidates for electron electric dipole moment searches [Verma, Jayich, and Vutha, Phys. Rev. Lett. 125, 153201 (2020)]. Here, we calculate potential energy curves, permanent electric dipole and quadrupole moments, and static electric dipole polarizabilities for the YbCu, YbAg, and YbAu molecules in their ground electronic states. We use the coupled cluster method restricted to single, double, and noniterative triple excitations with large Gaussian basis sets, while the scalar relativistic effects are included within the small-core energy-consistent pseudopotentials. We find that the studied molecules are relatively strongly bound with the well depths of 5708$\,$cm$^{-1}$, 5253$\,$cm$^{-1}$, 13349$\,$cm$^{-1}$ and equilibrium distances of 5.50$\,$bohr, 5.79$\,$bohr, 5.55$\,$bohr for the YbCu, YbAg, and YbAu, respectively. They have large permanent electric dipole moments of 3.2$\,$D, 3.3$\,$D, and 5.3$\,$D at equilibrium distances, respectively. We also calculate equilibrium geometries and energies of corresponding trimers. The studied molecules are chemically reactive unless they are segregated in an optical lattice or shielded with external fields. The investigated molecules may find application in ultracold controlled chemistry, dipolar many-body physics, or precision measurement experiments.   

\end{abstract}

\maketitle

\section{Introduction}

Ultracold polar molecules constitute excellent systems for studying the fundamentals of quantum physics and chemistry~\cite{CarrNJP09}. Rich and controllable internal molecular structure and intermolecular interactions allow for unique experiments on ultracold controlled chemistry~\cite{BohnScience17}, quantum simulations of many-body physics~\cite{GrossScience17}, and precision measurements~\cite{DeMilleScience17}. Highly accurate molecular spectroscopy can be employed to probe fundamental physics, including tests of fundamental symmetries~\cite{ChuppRMP19}, searches for the spatiotemporal variation of fundamental constants~\cite{KobayashiNC19}, measurements of the electric dipole moment of the electron~\cite{AndreevNature18}, the electron-to-proton mass ratio or the fine structure constant~\cite{AlighanbariNature20}, tests of quantum electrodynamics~\cite{AlighanbariNature20}, and others~\cite{SafronovaRMP18}.  

Recently, ultracold RaAg~\cite{SunagaPRA19} and YbAg~\cite{VermaPRL20} molecules in the $X^2\Sigma^+$ ground electronic state have been proposed as promising candidates for precision measurements and electron electric dipole moment searches. At the same time, molecules consisting of a Ag atom interacting with an alkali-metal or alkaline-earth-metal atom~\cite{SmialkowskiPRA21} have been shown to be strongly bound with highly polarized covalent or ionic bonds resulting in very large permanent electric dipole moments, significantly larger than in alkali-metal molecules. The RaAg molecule has been predicted to have the permanent electric dipole moment as large as 5.1$\,$D at the equilibrium distance and the potential well depth of 9563$\,$cm$^{-1}$~\cite{SmialkowskiPRA21}. To the best of our knowledge, the electronic structure calculations for the YbAg molecule and the analogous YbCu and YbAu molecules have not yet been reported in the literature. 

In this paper, to fill this gap and to facilitate the formation of such molecules for ultracold studies, we theoretically investigate the ground-state electronic properties of the YbCu, YbAg, and YbAu molecules. We compute potential energy curves, permanent electric dipole and quadrupole moments, and static electric dipole polarizabilities. We employ large Gaussian basis sets and the coupled cluster method restricted to single, double, and noniterative triple excitations to include the electron correlation. We use the small-core energy-consistent pseudopotentials to account for the scalar relativistic effects. We find that the studied molecules are relatively strongly bound and have large permanent dipole moments of 3.2$\,$D, 3.3$\,$D, and 5.3$\,$D at equilibrium distances for YbCu, YbAg and YbAu, respectively. The investigated molecules are chemically reactive unless segregated in an optical lattice or shielded with external fields.

The considered ultracold molecules can be formed from ultracold mixtures of closed-shell Yb and open-shell Cu, Ag, or Au atoms, following recent experimental advances in studies of Yb+Rb~\cite{NemitzPRA09}, Hg+Rb~\cite{WitkowskiOE17}, Sr+Rb~\cite{BarbeNP18}, Yb+Li~\cite{GreenPRX20}, and Yb+Cs~\cite{WilsonPRA21} combinations. Both photoassociation~\cite{JulienneRMP06a} and magnetoassociation~\cite{JulienneRMP06b} followed by the stimulated Raman adiabatic passage stabilization~\cite{VitanovRMP17} to the ground rovibrational level can potentially be employed. Bose-Einstein condensation~\cite{TakasuPRL03} and degenerate Fermi gases~\cite{FukuharaPRL07} of Yb have already been realized. Cu and Ag atoms have also been produced and trapped at ultralow temperatures using buffer-gas cooling and magnetic trapping~\cite{BrahmsPRL08} or magneto-optical cooling and trapping~\cite{UhlenbergPRA00}. Ultracold Au atoms may be obtained similarly but using UV lasers.

The structure of the paper is the following. In Section~\ref{sec:theory}, we describe the employed computational methods. In Section~\ref{sec:results}, we present and discuss the obtained results. In section~\ref{sec:summary}, we provide a summary and outlook.

\section{Computational methods}
\label{sec:theory}

We calculate potential energy curves in the Born-Oppenheimer approximation using the computational scheme recently applied to the ground electronic states of diatomic molecules consisting of a Cu or Ag atom interacting with an alkaline-earth-metal atom~\cite{SmialkowskiPRA21}. The interaction of a closed-shell Yb atom in the ground singlet $^1S$ electronic state with an open-shell Cu, Ag, or Au atom in the lowest doublet $^2S$ state results in the ground molecular electronic state of the doublet $X^2\Sigma^+$ symmetry.

The considered ground-state molecules are well described at all internuclear distances by single-reference methods. Therefore, we describe them with the spin-restricted open-shell coupled cluster method restricted to single, double, and non-iterative triple excitations (RCCSD(T))~\cite{PurvisJCP82,KnowlesJCP93}. The interaction energies $V(R)$, as functions of the internuclear distance $R$, are obtained with the supermolecular method with the basis set superposition error (BSSE) corrected by using the Boys-Bernardi counterpoise correction~\cite{BoysMP70},
\begin{equation}
V(R)=E_{AB}(R)-E_{A}(R)-E_{B}(R)\,,
\label{eq:intenergy}
\end{equation}
where $E_{AB}(R)$ is the total energy of the molecule $AB$, and $E_{A}(R)$ and $E_{B}(R)$ are the total energies of the atoms $A$ and $B$ computed in the diatom basis set, all at a distance $R$.

The topology of three-dimensional potential energy surfaces for the lowest $^2A'$, $^1A'$, and $^3A'$ electronic states of triatomic Yb$_2A$ and Yb$A_2$ molecules ($A$=Cu, Ag, Au) is studied using the second-order many-body (M{\o}ller-Plesset) perturbation theory~\cite{AmosCPL91}. Next, geometries and energies are optimized with the CCSD(T) method around global and local minima. BSSE is corrected by using the counterpoise correction. 

The scalar relativistic effects are included by employing the relativistic effective-core energy-consistent pseudopotentials (ECP) to replace the inner-shell electrons~\cite{DolgCR12}. The Cu, Ag, and Cu atoms are described with the ECP10MDF, ECP28MDF, and ECP60MDF pseudopotentials~\cite{FiggenCP05}, respectively, together with the aug-cc-pwCV5Z-PP basis sets designed for those ECPs~\cite{PetersonTCA05} ($i$ and $h$ exponents are omitted because of incompatibility with CPP). The Yb atom is described with the ECP60MDF effective-core pseudopotential together with the corresponding core-polarization potential (CPP)~\cite{WangTCA98} and the $[10s10p9d5f3g]$ basis set~\cite{SmialkowskiPRA20}. Thus, 10, 28, 60 and, 60 electrons in the inner shells are replaced by pseudopotentials, and remaining $3s^23p^63d^{10}4s^1$, $4s^24p^64d^{10}5s^1$, $5s^25p^65d^{10}6s^1$, and $5s^25p^66s^2$ electrons from Cu, Ag, Au, and Yb, respectively, are treated explicitly and correlated. To accelerate the convergence towards the complete basis set limit, the atomic basis sets are additionally augmented in all calculations for diatomic molecules by the set of the $[3s3p2d2f1g]$ bond functions~\footnote{Bond function exponents, $s$: 0.6, 0.2, 0.067, 0.02, $p$: 0.6, 0.2, 0.067, 0.02, $d$: 0.4, 0.13, 0.04 $f$: 0.4, 0.13, 0.04, $g$: 0.2.}. 

\begin{table}[tb!]
\caption{Characteristics of the Cu, Ag, Au, and Yb atoms: the static electric dipole polarizability $\alpha$, the ionization potential IP, the electron affinity EA, and the lowest $S$--$P$ excitation energy (${}^2S$--${}^2P$ for Cu, Ag, and Au and ${}^1S$--${}^3P$ for Yb). Present theoretical values are compared with the most accurate available experimental or theoretical data. Experimental excitation energies are averaged on spin-orbit manifolds. \label{tab:atoms}}  
\begin{ruledtabular}
\begin{tabular}{lllll}
Atom & $\alpha\,$(${e^2a_0^2}/{E_\text{h}}$)  & IP$\,$(cm$^{-1}$) & EA$\,$(cm$^{-1}$) & $S$--$P$$\,$(cm$^{-1}$)\\
\hline
Cu & 45.9 & 62406 & 10003 & 31062 \\
   & 46.5~\cite{NeogradyIJQC97} & 62317~\cite{nist} & 9967~\cite{BilodeauJPB98} & 30701~\cite{nist} \\
Ag & 50.2 & 61249 & 10608 & 30363 \\
   & 52.5~\cite{NeogradyIJQC97} & 61106~\cite{nist} & 10521~\cite{BilodeauJPB98} & 30166~\cite{nist} \\
Au & 36.3 & 74216 & 18519 & 40632  \\
   & 36.1~\cite{NeogradyIJQC97} & 74409~\cite{nist} & 18620~\cite{HotopJCP73} & 39903~\cite{nist} \\
Yb & 136.0 & 50479 & $\approx 0$ & 20119  \\
   & 139.3~\cite{BeloyPRA12} & 50443~\cite{nist} & $\approx 0$~\cite{AndersenJCP99} & 18869~\cite{nist} \\
\end{tabular}
\end{ruledtabular}
\end{table}

The long-range dispersion-interaction $C^\text{disp}_6=\frac{3}{\pi}\int_0^\infty \alpha_A(i\omega)\alpha_B(i\omega)d\omega$ coefficients are calculated from the atomic dynamic electric dipole polarizabilities at the imaginary frequency, $\alpha_{A(B)}(i\omega)$~\cite{JeziorskiCR94}. The dynamic polarizabilities of the Cu, Ag, and Au atoms are constructed as a sum over states using experimental energies~\cite{nist} and transition dipole moments from Refs.~\cite{TopcuPRA06,ZhangPRA08}. The dynamic polarizability of the Yb atom is obtained with the explicitly connected representation of the expectation value and polarization propagator within the coupled cluster method~\cite{KoronaMP06}. The long-range dispersion-interaction $C^\text{disp}_8$ coefficients are estimated by fitting the $-C^\text{disp}_6/R^6-C^\text{disp}_8/R^8$ formula with the calculated $C^\text{disp}_6$ coefficients to the calculated interaction potentials at interatomic distances between 12 and 30$\,$bohr.

The permanent electric dipole $d(R)$ and quadrupole $Q(R)$ moments and static electric dipole polarizabilities $\alpha(R)$ are calculated with the finite field approach. The $z$ axis is selected along the internuclear axis, oriented from the Cu, Ag, or Au atom to the Yb atom. The vibrationally averaged dipole moments $d_v$ are calculated as expectation values with radial vibrational wavefunctions.

To validate the accuracy of the employed electronic structure method and basis sets, we calculate the atomic properties such as the static electric dipole polarizabilities, the ionization potentials, the electron affinities, and the lowest electronic excitation energies for the considered atoms and compare them with the most accurate available experimental or theoretical data in Table~\ref{tab:atoms}. The very good agreement around 1\% for most properties suggests that high accuracy of molecular calculations with the present methods can also be expected. Based on the quality of atomic assessment, test molecular calculations in smaller basis sets, and our previous experience, we estimate the uncertainty of the present results to be of the order of 5\%.

All electronic structure calculations are performed with the \textsc{Molpro} package of \textit{ab initio} programs~\cite{Molpro,MOLPRO-WIREs}. Vibrational eigenstates $\varphi_v(R)$ and eigenenergies $E_v$ are computed using numerically exact diagonalization of the Hamiltonian for the nuclear motion within the discrete variable representation on the non-equidistant grid~\cite{TiesingaPRA98}. Atomic masses of the most abundant isotopes are assumed.

\section{Results and discussion}
\label{sec:results}

\subsection{Potential energy curves}

\begin{table*}[!tb]
\caption{Characteristics of the YbCu, YbAg, and YbAu molecules in the $X^2\Sigma^+$ ground electronic state: equilibrium interatomic distance~$R_e$, well depth~$D_e$, harmonic constant~$\omega_e$, rotational constant~$B_e$, permanent electric dipole moment~$d_e$, permanent electric quadrupole moment~$Q_e$, isotropic and anisotropic components of the static electric dipole polarizability~$\bar{\alpha}_e$ and $\Delta\alpha_e$, and number of vibrational levels~$N_v$.}
\label{tab:const} 
\begin{ruledtabular}
\begin{tabular}{lccccccccc}
Molecule & $R_e\,(a_0)$ & $D_e\,$(cm$^{-1}$) & $\omega_e\,$(cm$^{-1}$) & $B_e\,$(cm$^{-1}$) & $d_e\,$(D) & $Q_e\,$($ea_0^2$) & $\bar{\alpha}_e\,$($\frac{e^2a_0^2}{E_\text{h}}$) & $\Delta\alpha_e\,$($\frac{e^2a_0^2}{E_\text{h}}$) & $N_v$ \\
\hline
YbCu & 5.500 &  5708 & 144.3 & 0.0502 & 3.22 & -0.80 & 182 & 100 & 84 \\
YbAg & 5.788 &  5253 & 109.4 & 0.0271 & 3.30 &  0.51 & 191 & 121 & 105 \\
YbAu & 5.554 & 13349 & 125.0 & 0.0211 & 5.31 & -0.23 & 152 & 55 & 183 \\
\end{tabular}
\end{ruledtabular}
\end{table*}

\begin{figure}[tb]
\begin{center}
\includegraphics[width=\columnwidth]{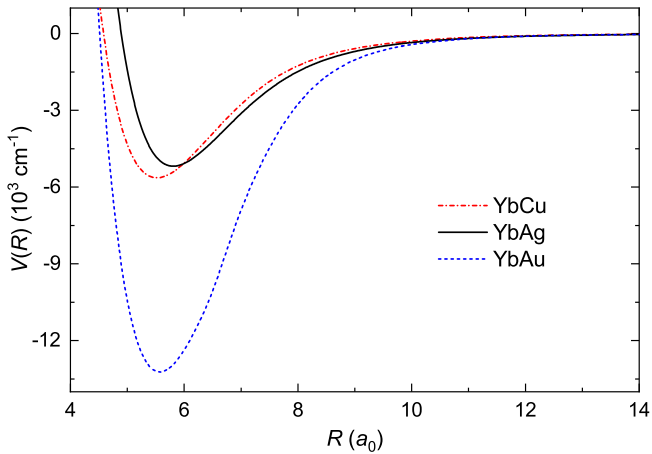}
\end{center}
\caption{Potential energy curves of the YbCu, YbAg, and YbAu molecules in the $X^2\Sigma^+$ ground electronic state.}
\label{fig:PES}
\end{figure}

The computed potential energy curves of the $X^2\Sigma^+$ symmetry for the YbCu, YbAg, and YbAu molecules are presented in Fig.~\ref{fig:PES}. The corresponding spectroscopic characteristics such as the equilibrium interatomic distance~$R_e$, well depth~$D_e$, harmonic constant~$\omega_e$, rotational constant~$B_e$, and number of vibrational levels~$N_v$ (for $j=0$) are collected in Table~\ref{tab:const}. 

All potential energy curves presented in Fig.~\ref{fig:PES} show a smooth behavior with well-defined minima. The potential energy curves of the YbCu and YbAg molecules are similar to each other, whereas the binding in the YbAu molecule is significantly stronger. This difference can be attributed to a significantly larger electronegativity (by the Pauling scale~\cite{Pauling1960}) of the Au (2.54) atom than that of the Cu (1.90) and Ag (1.93) atoms.  The YbCu and YbAg molecules have slightly smaller well depths, while the YbAu molecule has a slightly larger well depth than the molecules consisting of a Ag or Cu atom interacting with an alkaline-earth-metal atom~\cite{SmialkowskiPRA21}. All three studied molecules are significantly more strongly bound and have shorter equilibrium distances than alkali-metal--ytterbium~\cite{QinqinJPCA} and alkali-metal--alkaline-earth-metal molecules~\cite{PototschnigPCCP16}.  

\begin{table*}[tb!]
\caption{Long-range $C^\text{disp}_6$ and $C^\text{disp}_8$ coefficients for interatomic interactions and long-range $\tilde{C}_3^\text{dd}$, $\tilde{C}_4^\text{dq}$, $\tilde{C}^\text{disp}_6$, and $\tilde{C}_6^\text{rot}$ coefficients for intermolecular interactions.}
\label{tab:Cn} 
\begin{ruledtabular}
\begin{tabular}{lcccccc}
Molecule & $C^\text{disp}_6\,$($E_\mathrm{h}a^6_0$) & $C^\text{disp}_8\,$($E_\mathrm{h}a^8_0$) & $\tilde{C}_3^\text{dd}\,$($E_\mathrm{h}a^3_0$) & $\tilde{C}_4^\text{dq}\,$($E_\mathrm{h}a^4_0$) & $\tilde{C}^\text{disp}_6\,$($E_\mathrm{h}a^6_0$) &  $\tilde{C}_6^\text{rot}\,$($E_\mathrm{h}a^6_0$) \\
\hline
YbCu & 629 & $7.5\times 10^4$ & 1.60 & -1.01 & 3190 & $1.9\times 10^6$ \\
YbAg & 681 &  $11\times 10^4$ & 1.69 &  0.66 & 3429 & $3.8\times 10^6$ \\
YbAu & 563 &  $12\times 10^4$ & 4.36 & -0.48 & 2434 & $3.3\times 10^7$ \\
\end{tabular}
\end{ruledtabular}
\end{table*}

The relatively large binding energy and short equilibrium distances of the ground-state YbAu molecule may indicate the highly polarized covalent or even ionic nature of its chemical bond and significant stabilizing contribution of the electrostatic and induction interactions. The large difference in the electronegativity of the Au (2.54) and Yb (1.1) atoms may be responsible for a significant bond polarization and considerable contribution of the Yb$^+$Au$^-$ ionic configuration to its ground state bonding~\cite{Pauling1960}. This is further analyzed in the following subsection.  

The interaction potentials at large interatomic distances approach asymptotic behavior given by
\begin{equation}
V(R)\approx -\frac{C^\text{disp}_6}{R^6}-\frac{C^\text{disp}_8}{R^8}+\dots\,,
\end{equation}
where $C^\text{disp}_6$ and $C^\text{disp}_8$ are two leading long-range dispersion-interaction coefficients. The calculated coefficients for the studied systems are collected in Table~\ref{tab:Cn} and are smaller than for analogous alkali-metal and alkaline-earth-metal molecules because the polarizabilities of the Ag, Cu, and Au atoms are a few times smaller than the polarizabilities of alkali-metal and alkaline-earth-metal atoms.

\begin{figure}[!tb]
\begin{center}
\includegraphics[width=\columnwidth]{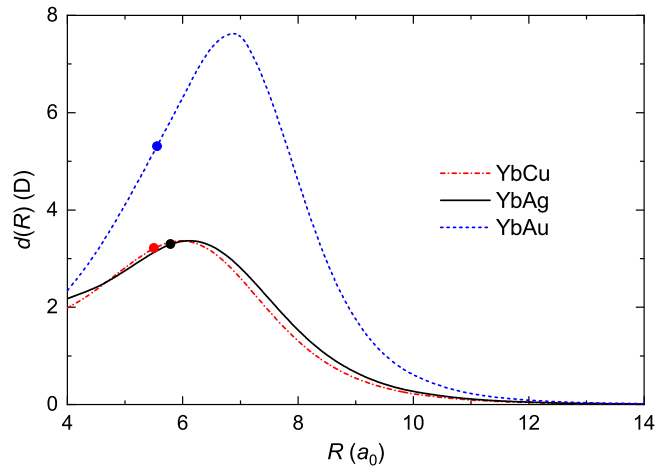}
\end{center}
\caption{Permanent electric dipole moments of the YbCu, YbAg, and YbAu molecules in the $X^2\Sigma^+$ ground electronic state. The dots indicate values for equilibrium distances.}
\label{fig:dip}
\end{figure}

\subsection{Permanent electric dipole and quadrupole moments}

Permanent electric dipole moments $d(R)$ as functions of the interatomic distance for the YbCu, YbAg, and YbAu molecules in the $X^2\Sigma^+$ ground electronic states are presented in Fig.~\ref{fig:dip}. The corresponding values at equilibrium distances $d_e\equiv d(R_e)$ are collected in Table~\ref{tab:const}.

The permanent electric dipole moment curves of the YbCu and YbAg molecules are similar to each other (similarly to potential energy curves), whereas the charge polarization in the YbAu molecule is significantly stronger. Again, this difference can be attributed to a significantly larger electronegativity of the Au atom than that of the Cu and Ag atoms. Permanent electric dipole moments can be used to measure the bond polarization and ionic character $IC$ of the studied molecules, e.g., by calculating the ratio of the permanent electric dipole moment of a given molecule, $d_e$, at the equilibrium distance, $R_e$, to the maximal possible value, $d_\text{max}=eR_e$, corresponding to a purely ionic molecule~\cite{Pauling1960},
\begin{equation}
IC=\frac{d_{e}}{d_\text{max}}=\frac{d_{e}}{eR_{e}}\,.\\
\end{equation}
The calculated ratios are 23\%, 22\%, and 38\% for YbCu, YbAg, and YbAu, respectively. These values agree with Mulliken and natural orbital population analysis~\cite{ReedJCP85}. Thus, the bonds in the YbCu and YbAg molecules are considerably polarized, whereas, in the YbAu molecule, the admixture of the Yb$^+$Au$^-$ ionic configuration in the ground state may be significant. The observed trends coincide with the differences in atomic electronegativities $\chi$ of the Cu (1.90), Ag (1.93), Au (2.54), and Yb (1.1) atoms. The calculated ratios also agree with the approximate percent ionic character of a heteronuclear diatomic molecule with a single bond given by Pauling~\cite{Pauling1960} $1-\exp[-(\chi_\text{Cu/Ag/Au}-\chi_\text{Yb})]$, resulting in 18\%, 19\%, and 30\% for YbCu, YbAg, and YbAu, respectively. The permanent electric dipole moment of YbAu increases linearly with the interatomic distance in the vicinity of the interaction potential well, additionally confirming its partially ionic character.

The YbCu, YbAg, and YbAu molecules have permanent electric dipole moments significantly larger than alkali-metal--ytterbium~\cite{QinqinJPCA} and alkali-metal--alkaline-earth-metal molecules~\cite{PototschnigPCCP16}, but similar to molecules consisting of a Ag or Cu atom interacting with an alkaline-earth-metal atom~\cite{SmialkowskiPRA21}. The YbAu molecule has the permanent electric dipole moment as large as the most polar alkali-metal LiCs molecule~\cite{AymarJCP05}.

\begin{figure}[!tb]
\begin{center}
\includegraphics[width=\columnwidth]{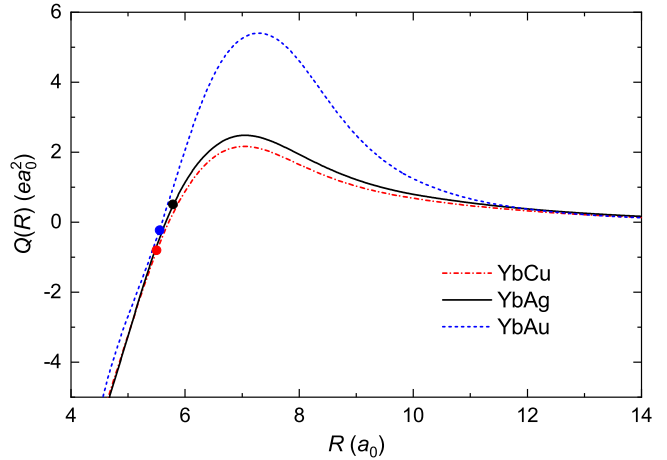}
\end{center}
\caption{Permanent electric quadrupole moments of the YbCu, YbAg, and YbAu molecules in the $X^2\Sigma^+$ ground electronic state. The dots indicate values for equilibrium distances.}
\label{fig:quad}
\end{figure}

The permanent electric quadrupole moments $Q(R)$ as functions of the interatomic distance for the YbCu, YbAg, and YbAu molecules in the $X^2\Sigma^+$ ground electronic states are presented in Fig.~\ref{fig:quad}. Again, curves for the YbCu and YbAg molecules are similar to each other. The corresponding values at equilibrium distances $Q_e\equiv Q(R_e)$ are collected in Table~\ref{tab:const}. The calculated quadrupole moments are relatively small, especially around equilibrium distances, where quadrupole moment curves change the sign.

\begin{figure}[!tb]
\begin{center}
\includegraphics[width=\columnwidth]{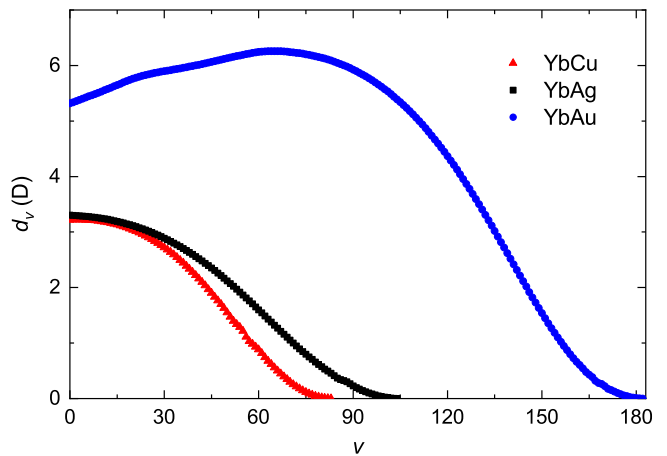}
\end{center}
\caption{Permanent electric dipole moments of the YbCu, YbAg, and YbAu molecules in different vibrational levels of the $X^2\Sigma^+$ ground electronic state as a function of the vibrational quantum number.}
\label{fig:dv}
\end{figure}

The permanent electric dipole moments of the YbCu, YbAg, and YbAu molecules in different vibrational levels of their ground electronic state $d_v=\int |\varphi_v(R)|^2d(R)dR$ as a function of the vibrational quantum number $v$ are presented in Fig.~\ref{fig:dv}. For YbCu, the largest value is 3.22$\,$D in the level with $v=3$ and $E_b=-5208\,$cm$^{-1}$. For YbAg, the largest value is 3.30$\,$D in the level with $v=0$ and binding energy $E_b=-5199\,$cm$^{-1}$. For YbAu, the largest value is 6.26$\,$D in the level with $v=65$ and $E_b=-5756\,$cm$^{-1}$, while it is 5.33$\,$D in the level with $v=0$ and $E_b=-13287\,$cm$^{-1}$. The increase of the permanent electric dipole moment with increasing the vibrational quantum number and decreasing the vibrational binding energy is visible for the YbAu molecule with $v<65$ due to the observed increase of its permanent electric dipole moment with the interatomic distance (c.f.~Fig.~\ref{fig:dip}). Thus, large permanent electric dipole moments for YbAu molecules in highly excited vibrational levels may allow for new molecular control schemes.

\begin{table}[bt!]
\caption{Characteristics of dipolar molecules and their intermolecular interactions: ground-state permanent electric dipole moment $d_e$, polarizing electric field $\mathcal{E}_\mathrm{pol}$, characteristic length of dipolar interaction $a_\text{dd}$, and characteristic nearest-neighbor energy shift  $V_\mathrm{dd}=C_3^\text{dd}/(\lambda/2)^3$ for molecules in an optical lattice formed by $\lambda=$1064$\,$nm laser. Results for the YbCu, YbAg, and YbAu molecules are compared with parameters for other molecules used in ultracold experiments~\cite{GadwayJPB16}.}
\label{tab:dip} 
\begin{ruledtabular}
\begin{tabular}{lcccc}
Molecule & $d_e$(D) & $\mathcal{E}_\mathrm{pol}\,$(V/cm) & $a_\mathrm{dd}\,$($10^3\,a_0$) & $V_\mathrm{dd}\,$(kHz)  \\
\hline
YbCu & 3.22 & 1860 & 231 &  10.4\\
YbAg & 3.30 & 978 & 288 & 10.9 \\
YbAu & 5.31 & 474 & 982 & 28.2 \\
CsAg~\cite{SmialkowskiPRA21} & 9.75 & 329 & 2144 & 95.3 \\
KRb~\cite{NiScience08} & 0.57 & 7832 & 4 & 0.3  \\
NaRb~\cite{GuoPRL16} & 3.2 & 2594 & 106 & 10.3 \\
LiCs~\cite{DeiglmayrPRL08} & 5.5 & 4071 & 398 & 30.3 \\
RbSr~\cite{CiameiPCCP18} & 1.5 & 1467 & 37 & 2.3 \\
CaF~\cite{TruppeNP17} & 3.1 & 13287 &  52 & 9.4 \\
\end{tabular}
\end{ruledtabular}
\end{table}

The relatively large permanent electric dipole moments combined with large reduced masses and small rotational constants of the investigated molecules open the way for their applications in quantum simulations of strongly interacting dipolar quantum many-body systems, controlled chemistry, and precision measurements. Such applications often require molecular polarization with the external static electric field. The characteristic scale of the electric field needed to polarize molecules can be quantified by
\begin{equation}
\mathcal{E}_\mathrm{pol}=\frac{2B_e}{d_e}\,.
\end{equation}
Values of the polarizing electric field for the YbCu, YbAg, and YbAu molecules are presented in Table~\ref{tab:dip}, together with values for other molecules relevant for ongoing ultracold experiments. For the YbAg molecule, $\mathcal{E}_\mathrm{pol}$ is below 1$\,$kV/cm, which is smaller and more favorable than for most of the other ultracold molecules. 

The intermolecular interactions between the studied molecule at ultralow temperatures are dominated by the long-range dipolar interaction, which for the polarized molecules takes the form
\begin{equation}
V_\text{dd}(R,\theta)=\frac{\tilde{C}^\text{dd}_3(1-3\cos^2\theta)}{R^3}\,,
\end{equation}
where $\theta$ is the angle between the directions of polarization and intermolecular axis, and the long-range dipole-dipole electrostatic-interaction coefficient $\tilde{C}^\text{dd}_3$ is given by
\begin{equation}
\tilde{C}^\text{dd}_3=\frac{d_v^2}{4\pi\varepsilon_0}\,,
\end{equation}
where $d_v$ is the permanent electric dipole moment of molecules in a state $v$. The $\tilde{C}^\text{dd}_3$ coefficients for the studied ground-state molecules are collected in Table~\ref{tab:Cn}.

The dipole-dipole interactions can be further characterized by an effective dipolar length $a_\text{dd}$~\cite{LahayeCPP09} defined by
\begin{equation}
a_\text{dd}=\frac{d_v^2m}{12\pi\varepsilon_0\hbar^2}\,,
\end{equation}
where $m$ is the molecule mass. The dipolar lengths for the studied ground-state molecules are collected in Table~\ref{tab:dip} and compared with values for other molecules relevant for ongoing ultracold experiments. They are significantly larger for the present molecules than for other $^2\Sigma$-state molecules and comparable to the most dipolar alkali-metal molecules.  

Another important parameter characterizing dipole-dipole interactions is the energy shift between nearest-neighbor molecules in an optical lattice  $V_\mathrm{dd}=C_3^\text{dd}/(\lambda/2)^3$, where $\lambda$ is the laser wavelength. Its values for the studied ground-state molecules are collected in Table~\ref{tab:dip} and compared with values for other molecules relevant for ongoing ultracold experiments. Again, they are comparable or larger for the present molecules than for other molecules.

Higher-order electrostatic terms of intermolecular interactions with more complex orientation dependences include the dipole-quadrupole $\sim{d_vQ_v}/{R^4}$, quadrupole-quadrupole $\sim{Q_v^2}/{R^5}$, and dipole-octupole $\sim{d_vO_v}/{R^5}$ contributions~\cite{Stone2013}. The long-range dipole-quadrupole electrostatic-interaction coefficients,
\begin{equation}
\tilde{C}^\text{dq}_4=\frac{d_vQ_v}{4\pi\varepsilon_0}\,,
\end{equation}
for the studied ground-state molecules are also collected in Table~\ref{tab:Cn}, but have small valued because of small molecular quadrupole moments. 

If molecules are not polarized by an external electric field, then in their ground rotational states, their interaction is dominated by  the effective isotropic term $-\tilde{C}_6^\text{rot}/R^6$, resulting from the dipolar interaction in the second-order of perturbation theory and given by the long-range coefficient 
\begin{equation}
\tilde{C}_6^\text{rot}=\frac{d_v^4}{6B_v}\,,
\end{equation}
where $B_v$ is the rotational constant for a vibrational state $v$. The $\tilde{C}_6^\text{rot}$ coefficients for the studied ground-state molecules are collected in Table~\ref{tab:Cn}.

\subsection{Static electric dipole polarizabilities}

\begin{figure}[tb]
\begin{center}
\includegraphics[width=\columnwidth]{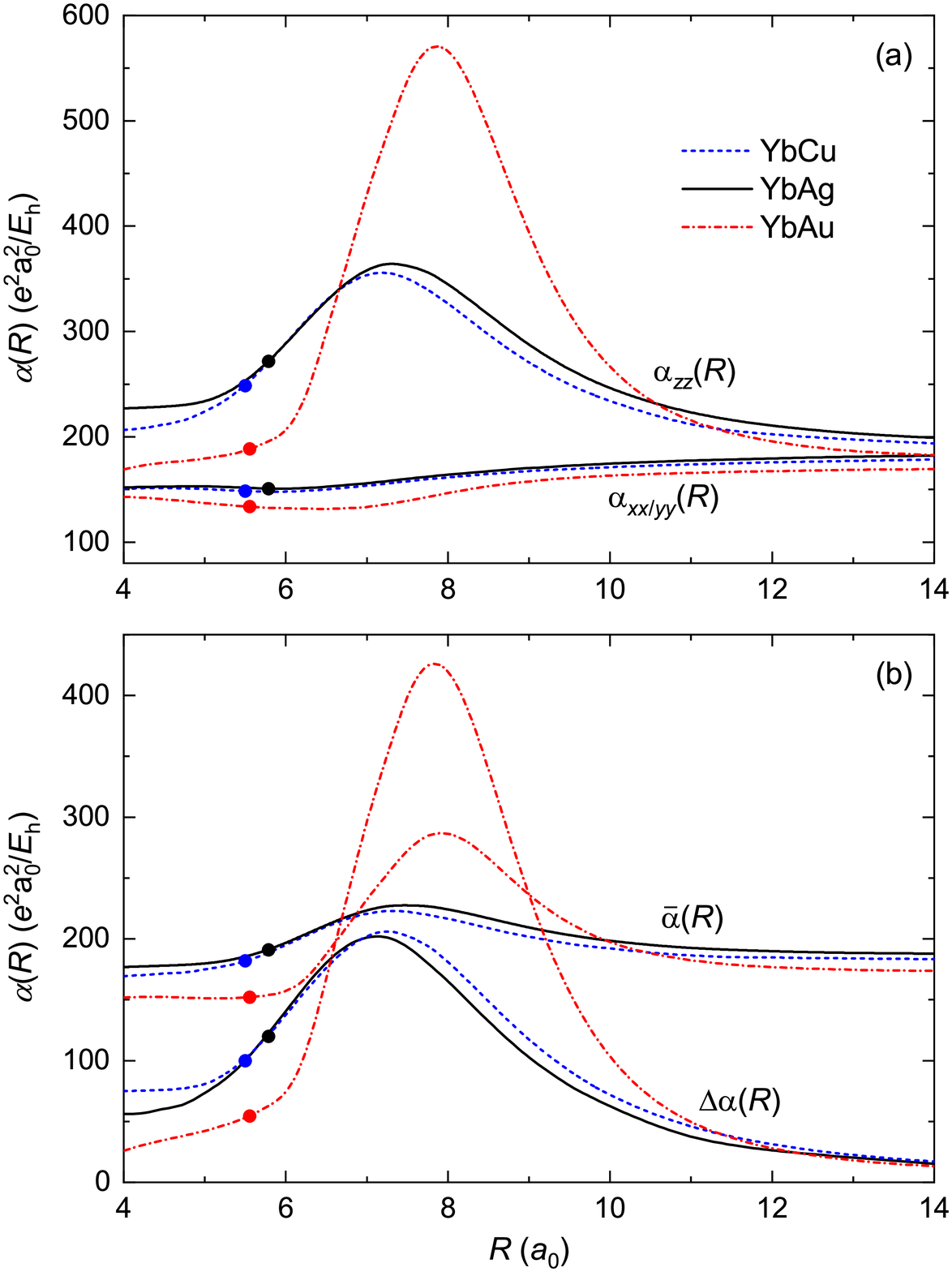}
\end{center}
\caption{Static electric dipole polarizabilities of the YbCu, YbAg, and YbAu molecules in the $X^2\Sigma^+$ ground electronic state: (a) parallel and perpendicular and (b) isotropic and anisotropic components. The dots indicate values for equilibrium distances.}
\label{fig:pol}
\end{figure}

The computed parallel $\alpha^\parallel(R)\equiv\alpha^{zz}(R)$ and perpendicular $\alpha^\perp(R)\equiv\alpha^{xx}(R)=\alpha^{yy}(R)$ components of the static electric dipole polarizability tensor as functions of the interatomic distance for the YbCu, YbAg, and YbAu molecules in the $X^2\Sigma^+$ electronic states are presented in Fig.~\ref{fig:pol}(a). The polarizabilities for YbCu and YbAg are similar to each other, while the interaction-induced variation for YbAu is much more pronounced. At large interatomic distances, they approach their asymptotic behavior given by the atomic polarizabilities $\alpha_A$ and $\alpha_B$~\cite{JensenJCP02}
\begin{equation}
\begin{split}
\alpha^\parallel(R)&\approx \alpha_A+\alpha_B+\frac{4\alpha_A\alpha_B}{R^3}+\frac{4(\alpha_A+\alpha_B)\alpha_A\alpha_B}{R^6}\,,\\
\alpha^\perp(R)&\approx \alpha_A+\alpha_B-\frac{2\alpha_A\alpha_B}{R^3}+\frac{(\alpha_A+\alpha_B)\alpha_A\alpha_B}{R^6}\,.
\end{split}
\end{equation}

The isotropic $\bar{\alpha}(R)$ and anisotropic $\Delta\alpha(R)$ components of the static electric dipole polarizability can also be obtained from $\alpha^\perp(R)$ and $\alpha^\parallel(R)$
\begin{equation}
\begin{split}
\bar{\alpha}(R)&=\frac{2\alpha^\perp(R)+\alpha^\parallel(R)}{3} \,,\\
\Delta\alpha(R)&=\alpha^\parallel(R)-\alpha^\perp(R)\,.
\end{split}
\end{equation}
They are presented in Fig.~\ref{fig:pol}(b). Their values at the equilibrium distances, $\bar{\alpha}_e\equiv\bar{\alpha}(R_e)$ and $\Delta\alpha_e\equiv\Delta\alpha(R_e)$, are collected for the studied molecules in Table~\ref{tab:const}. The equilibrium values are relatively close to the asymptotic ones, despite a large variation of the calculated polarizabilities at intermediate distances, especially for the YbAu molecule.

The polarizability describes the molecular response to the electric field in the second order of perturbation theory. For example, optical dipole trapping is governed by the isotropic polarizability $\bar{\alpha}$, while the laser-induced molecular alignment is controlled by the anisotropy of the polarizability $\Delta\alpha$~\cite{LemeshkoMP13}. Molecular polarizabilities may also be useful in the evaluation of intermolecular interactions~\cite{JeziorskiCR94}.

The leading long-range dispersion-interaction coefficients can be estimated by combination rules, e.g.~by the Slater-Kirkwood approximation~\cite{SlaterPR31}, 
\begin{equation}\label{eq:C6ij}
C^\text{disp}_{6,ij}\approx\frac{2}{3}\frac{\alpha_i\alpha_j}{(\alpha_i/N_i)^{1/2}+(\alpha_j/N_j)^{1/2}}\,,
\end{equation}
where $\alpha_{i(j)}$ is the static electric dipole polarizability of the $i(j)$ monomer and $N_{i(j)}$ represents its effective number of electrons. The effective number of electrons can be roughly approximated by the number of valence electrons or calculated from the known coefficients between like monomers, resulting in another known expression~\cite{KramerJCP70} 
\begin{equation}\label{eq:C6ii}
C^\text{disp}_{6,ij}\approx\frac{2 \alpha_i \alpha_j C^\text{disp}_{6,ii}C^\text{disp}_{6,jj}}{\alpha_i^2 C^\text{disp}_{6,jj}+\alpha_i^2 C^\text{disp}_{6,ii}}\,.
\end{equation}

Both Eqs.~\eqref{eq:C6ij} and \eqref{eq:C6ii} reproduce exact interatomic $C_6^\text{disp}$ coefficients for YbAg, YbCu, and YbAu with accuracy better than 10\%. Therefore, we calculate the intermolecular long-range dispersion-interaction coefficients $\tilde{C}^\text{disp}_6$ between ground-state molecules using Eq.~\eqref{eq:C6ij} with the isotropic polarizabilities $\bar{\alpha}_e$ and the effective number of electrons $N=3$ and collect them in Table~\ref{tab:Cn}. The values for the YbCu and YbAg molecules agree within 10\% with the rough approximation neglecting intramolecular interactions, $\tilde{C}^\text{disp}_6\approx 2C^\text{disp}_{6,\text{Yb}A}+C^\text{disp}_{6,A_2}+C^\text{disp}_{6,\text{Yb}_2}$ (the agreement for the YbAu molecule is worse because of its highly polarized bond). The calculated intermolecular electronic dispersion $\tilde{C}^\text{disp}_6$ coefficients are, however, negligibly small as compared to the rotational $\tilde{C}^\text{rot}_6$ ones (cf.~Tabel~\ref{tab:Cn}).

\subsection{Chemical reactions}

\begin{table}[bt!]
\caption{Characteristics of the triatomic Yb$_2A$ and Yb$A_2$ molecules ($A=$ Cu, Ag, Au): electronic state symmetry, equilibrium angle formed by $ABC$ atoms $\theta_e^{ABC}$, equilibrium distance between $A$ and $B$ atoms $R_e^{AB}$,  equilibrium distance between $B$ and $C$ atoms $R_e^{BC}$, and well depth $D_e$.}
\label{tab:3b} 
\begin{ruledtabular}
\begin{tabular}{lccccc}
$ABC$ & Symm. & $\theta_e^{ABC}$(deg) & $R_e^{AB}\,$($a_0$) & $R_e^{BC}\,$($a_0$)  & $D_e\,$(cm$^{-1}$)   \\
\hline
YbCuYb & $^2\Sigma^+_g$ & 180 & 5.79 & 5.79 & 8792 \\
YbCuCu & $^1\Sigma^+$   & 180 & 5.68 & 4.26 & 20217 \\
CuYbCu & $^1\Sigma^+_g$ & 180 & 5.43 & 5.43 & 19213 \\
CuYbCu & $^3B_2$  & 45.7 & 5.70 & 5.70 & 15850 \\
\hline
YbAgYb & $^2\Sigma^+_g$ & 180 & 6.04 & 6.04 & 8417 \\
YbAgAg & $^1\Sigma^+$   & 180 & 5.92 & 4.84 & 17890 \\
AgYbAg & $^1\Sigma^+_g$ & 180 & 5.71 & 5.71 & 18034 \\
AgYbAg & $^3B_2$        & 50.7 & 5.97 & 5.97 & 13873 \\
\hline
YbAuYb & $^2\Sigma^+_g$ &  180 & 6.00 & 6.00 & 14757 \\
YbAuAu & $^1\Sigma^+$   &  180 & 5.58 & 4.80 & 27857 \\
AuYbAu & $^1\Sigma^+_g$ &  180 & 5.53 & 5.53 & 33913 \\
AuYbAu & $^3B_2$        & 50.6 & 5.84 & 5.84 & 26041 \\
\end{tabular}
\end{ruledtabular}
\end{table}

The stability of the studied molecules against chemical reactions may be assessed using the calculated potential well depths and related dissociation energies~\cite{ZuchowskiPRA10,TomzaPRA13a,SmialkowskiPRA20}. Among the investigated species, the most-strongly-bound YbAu molecules in the rovibrational ground state of the $X^2\Sigma^+$ ground electronic state are chemically stable against atom-exchange reactions, i.e,
\begin{equation}\label{eq:YbAu}
2\,\text{YbAu}(X^{2}\Sigma^+) \not\to \text{Au}_2(X^{1}\Sigma^+_g) + \text{Yb}_2(X^{1}\Sigma^+_g)\,.
\end{equation}
On the other hand, two other molecules are chemically reactive
\begin{equation}
\begin{split}
2\,\text{YbCu}(X^{2}\Sigma^+) & \to \text{Cu}_2(X^{1}\Sigma^+_g) + \text{Yb}_2(X^{1}\Sigma^+_g)\,,\\
2\,\text{YbAg}(X^{2}\Sigma^+) & \to \text{Ag}_2(X^{1}\Sigma^+_g) + \text{Yb}_2(X^{1}\Sigma^+_g)\,,\\
\end{split}
\end{equation}
because the binding of the Cu$_2$ and Ag$_2$ dimers~\cite{SmialkowskiPRA21} is much stronger than that of the YbCu and YbAg molecules.    

The chemical reactivity of the YbCu and YbAg molecules may potentially be suppressed by spin-polarizing molecules with an external magnetic field, restricting the collision dynamics to high-spin intermolecular interaction potentials~\cite{TomzaPRA13a}. In the triplet state, the following atom-exchange reactions are energetically forbidden
\begin{equation}
\begin{split}\label{eq:YbCu}
2\,\text{YbCu}(X^{2}\Sigma^+) & \not\to \text{Cu}_2(a^{3}\Sigma^+_u) + \text{Yb}_2(X^{1}\Sigma^+_g)\,,\\
2\,\text{YbAg}(X^{2}\Sigma^+) & \not\to \text{Ag}_2(a^{3}\Sigma^+_u) + \text{Yb}_2(X^{1}\Sigma^+_g)\,.\\
\end{split}
\end{equation}
Unfortunately, the spin-relaxation mediated by the magnetic spin-spin and second-order spin-orbit coupling may lead to reactive singlet intermolecular interaction potentials~\cite{JanssenPRL13,Hermsmeier2021}. 

Except for the atom-exchange reactions, the trimer formation reactions may be another path of chemical losses~\cite{ZuchowskiPRA10,TomzaPRA13a,SmialkowskiPRA20}:
\begin{subequations}
\begin{eqnarray}\label{eq:thrimers}
2\,\text{Yb}A(X^{2}\Sigma^+) & \to \text{Yb}_2A(X^2A')  + A(^2S)\,,\label{eq:tria}\\
2\,\text{Yb}A(X^{2}\Sigma^+) & \to \text{Yb}A_2(X^1A')  + \text{Yb}(^1S)\,,\label{eq:trib}\\
2\,\text{Yb}A(X^{2}\Sigma^+) & \to \text{Yb}A_2(a^3A')  + \text{Yb}(^1S)\label{eq:tric}\,,
\end{eqnarray}
\end{subequations}
with $A$=Cu, Ag, or Au. Depending on the topology of potential energy surfaces, the $^1A'$ and $^1A'$ electronic states coreduce to $^1A_1$ and $^2A_1$ for isosceles triangular equilibrium geometries or $^1\Sigma^+$ and $^2\Sigma^+$ for linear ones, whereas the $^3A'$ electronic state coreduce to $^3B_2$ for isosceles triangular equilibrium geometry. Results of electronic structure calculations for triatomic Yb$_2A$ and Yb$A_2$ molecules are collected in Table~\ref{tab:3b}. We find that the reactions~\eqref{eq:tria} leading to the linear Yb$A$Yb($^2\Sigma^+_g$) molecules are energetically suppressed for all the studied diatomic molecules in their ground states, while the reactions~\eqref{eq:trib} leading to the strongly bound linear $A$Yb$A$($^1\Sigma^+_g$) or Yb$AA$($^1\Sigma^+$) molecules are exothermic for all the studied diatomic molecules. The reactions~\eqref{eq:tric} leading to the isoscales triangular $A_2\text{Yb}(^3B_2)$ molecules are exothermic for YbCu and YbAg, while nearly thermoneutral for YbAu. The binding energies of trimers are large, mostly more than twice larger than that of Yb$A$($X^{2}\Sigma^+$) dimers, because of stabilizing electrostatic and three-body interactions.

There are no reaction barriers for all the considered above chemical reactions, and there are exothermic reactive channels for all the investigated molecules, therefore all of them are chemically reactive even at ultralow temperatures. This opens the way for studying ultracold controlled chemical reactions. If chemical reactions are not desired, e.g.~in precision measurements or quantum many-body simulations, then the molecules should be protected from binary collisions by segregation in an optical lattice or shielding with external electromagnetic fields. Shielding with a microwave field~\cite{Anderegg2021} or by polarizing with an external electric field in a reduced dimensionality~\cite{MatsudaScience20} has already been experimentally demonstrated.

\section{Summary and conclusions}
\label{sec:summary}

Motivated by the experimental progress on formation and application of ultracold Yb+Rb~\cite{NemitzPRA09}, Hg+Rb~\cite{WitkowskiOE17}, Sr+Rb~\cite{BarbeNP18}, Yb+Li~\cite{GreenPRX20}, and Yb+Cs~\cite{WilsonPRA21} mixtures and recent theoretical proposal for using ultracold YbAg molecules for electron electric dipole moment searches~\cite{VermaPRL20}, we have studied electronic properties of the YbCu, YbAg, and YbAu molecules. We have calculated potential energy curves, permanent electric dipole and quadrupole moments, and static electric dipole polarizabilities using the coupled cluster method restricted to single, double, and noniterative triple excitations with large Gaussian basis sets and small-core energy-consistent pseudopotentials.

We have found that the studied molecules are relatively strongly bound and have relatively large permanent electric dipole moments. For YbAu, the maximal electric dipole moment exceeds 6.2$\,$D for highly excited vibrational level. We have also assessed possible channels of chemical reactions based on the energetics of the reactants and products and found that the considered molecules are chemically reactive. The investigated molecules may find application in ultracold controlled chemistry, dipolar many-body physics, or precision measurement experiments.

\begin{acknowledgments}
Financial support from the National Science Centre Poland (Grant No.~2016/23/B/ST4/03231) and the Foundation for Polish Science within the First Team program co-financed by the European Union under the European Regional Development Fund is gratefully acknowledged. The computational part of this research has been partially supported by the PL-Grid Infrastructure.
\end{acknowledgments}

\bibliography{YbAg}

\end{document}